# Anisotropic hot carrier relaxation mediated by electron–phonon scattering in TiN thin films

Hemant Verma[a,b,c,⊥], Tzu-Yu Peng[b,⊥], Shyr-Shyan Yeh[b], Sheng-Chieh Huang[b], Pritam Sardar[a,b,c], Yang-Hao Chan[d,*], Yu-Jung Lu [a,b,c,*], and Chao-Cheng Kaun[b,c,*]

[a] *Department of Physics, National Taiwan University, Taipei 10617, Taiwan*

[b] *Research Center for Applied Sciences, Academia Sinica, Taipei 11529, Taiwan*

[c] *Nano Science and Technology Program, TIGP, Academia Sinica, Taipei 11529, Taiwan*

[d] *Institute of Atomic and Molecular Sciences, Academia Sinica, Taipei 10617, Taiwan*

[⊥] *These authors contributed equally to this work.*

## ABSTRACT

Crystal orientations can shape the ultrafast energy relaxations of transition-metal nitride thin films. Here, we investigate the orientation-dependent electron-phonon (e-ph) mediated relaxation in titanium nitride (TiN) thin films along the [100], [110], and [111] directions by combining first-principles calculations with ultrafast pump-probe transient absorption spectroscopy. Using maximally localized Wannier functions, we evaluate e-ph quasiparticle scattering lifetimes near the Fermi level and identify a clear anisotropy: The TiN [111] orientation exhibits a longer e-ph scattering lifetime (15.96 fs) than [100] (13.69 fs) and [110] (11.12 fs), indicating reduced intrinsic e-ph scattering strength. Furthermore, we grew quasi-epitaxial, orientation-controlled TiN thin films on MgO substrates. Pump-probe measurements reveals that the population-level relaxation (hot-electron cooling) time also depends on orientations, with [111] films showing a significantly slower decay (110 fs) than [100] (90 fs) and [110] (80 fs). We emphasize that the calculated few-femtosecond scattering lifetimes and the measured few-hundred-femtosecond cooling time respectively represent single-event scattering and collective cooling, yet they exhibit consistent trends. These results demonstrate that crystallographic orientation provides a practical and powerful route to tune e-ph-governed relaxation in TiN thin films, offering essential design guidelines for refractory plasmonic and energy-conversion platforms.

*Corresponding authors. *E-mail addresses*: kauncc@gate.sinica.edu.tw (C.-C. Kaun), yujunglu@gate.sinica.edu.tw (Y.-J. Lu), yanghao@gate.sinica.edu.tw (Y.-H. Chan)

## 1. INTRODUCTION.

The investigation of hot carrier dynamics in transition metal nitrides (TMNs) has been emerging as a pivotal area of research, with significant implications for the development of next-generation energy devices [1-4]. As the need for efficient energy conversion technologies intensifies, TMNs, particularly titanium nitride (TiN), have garnered attention due to their exceptional electronic properties, thermal stability, and robust performance under extreme temperature conditions [5-7]. The unique characteristics of TMNs allow them to function as viable alternatives to conventional noble metals, such as gold and silver, particularly in applications where enhanced hot carrier dynamics are critical [8-12]. Moreover, in thin-film architectures, these dynamics are inherently surface-sensitive, making the choice of substrate and crystallographic orientation important levers for controlling energy dissipation pathways [13].

Electron–phonon (e–ph) interactions are key in transferring energy between electrons and lattice vibrations, significantly influencing hot carrier dynamics and transport properties. In the hot carrier regime, e–ph scattering predominantly governs carrier relaxation after the initially photoexcited non-thermal carrier population rapidly evolves via carrier–carrier processes (e.g., electron–electron scattering and momentum randomization). Thus, the experimentally observed pump–probe recovery in metallic TiN is more appropriately interpreted as hot-electron cooling (energy relaxation to the lattice), rather than semiconductor-like interband recombination [14-16]. However, ultrafast effects are prominently observed in the coherent and non-thermal regimes, driven by carrier–carrier scattering, momentum scattering, inter-valley scattering, or the confinement of carriers in quantum wells. Theoretically, A. Habib et al. recently reported using an e–ph interactions-based approach to explain the hot carrier dynamics in the TMNs to compare group IV and V with their various crystal structures. They conclude rock-salt TiN and hexagonal vanadium nitride (VN) as the most suitable candidates for large hot carrier mean free paths [17]. Following a similar approach, C. Jian et al. reported another comparative study of their theoretical stable configurations of rock-salt TiN and hexagonal VN at different temperatures, observing a significant effect of temperature on the hot carrier lifetimes and mean free paths [18]. Furthermore, J. Zhou et al. implemented another model to capture the impact of e–ph interactions on phonon transport, where they backed up their experimental observations of pump-probe spectroscopy analysis with a theoretical approach to e–ph interactions, utilizing ab initio simulations [19]. C. O. Karaman et al. reported the hot carrier dynamics in ultrathin monocrystalline gold, theoretically exemplified by e–ph interaction models [20]. These studies bridge ultrafast experiments with first-principles descriptions of energy flow, particularly in metallic systems where cooling is governed by coupled electron–phonon and phonon–phonon pathways [21]. Recent advancements in theoretical modeling, including two-temperature models and ab initio simulations, have yielded profound insights into the behavior of hot carriers within these nitride materials [22]. Nevertheless, how hot-carrier relaxation and energy dissipation evolve across different crystallographic orientations, especially in epitaxial TiN films, remains rarely explored [17, 18].

In this work, we investigate the hot carrier dynamics in TiN along various crystal orientations from first principles and ultrafast pump–probe spectroscopy measurements. Our study demonstrates the anisotropic nature of hot carriers across the [100], [110], and [111] crystallographic directions, revealing how crystal orientations profoundly influence the electronic structures and the lattice-mediated energy dissipation pathways that determine the measured cooling dynamics. Furthermore, we elucidate the mechanisms underpinning thermalization processes (e–ph interactions) that govern the relaxation dynamics of photo-excited carriers under non-equilibrium conditions at room temperature. Our results not only enhance an understanding of hot carrier dynamics in TiN crystal orientations, but also provide designing insights for energy conversion applications.

## 2. METHODS.

***2.1 Computational details:*** To efficiently calculate e–ph properties in a slab model, the JDFTx code [23] based on density functional theory (DFT) with SG15 norm-conserving pseudopotentials [24], within the generalized gradient approximation in the Perdew–Burke–Ernzerhof form [25]. A kinetic cutoff energy of 38 $E_h$ (Hartree energy) and a Fermi–Dirac smearing of 0.01 $E_h$ were utilized in our calculations. For Brillouin zone sampling, a $16 \times 16 \times 1$ k-grid was utilized, and to prevent artificial interactions in our surface-oriented calculations, a 15 Å vacuum layer was incorporated. All geometric structures were fully relaxed until the electronic energy and force components were less than $1 \times 10^{-8}$ eV and $1 \times 10^{-2}$ eV/Å, respectively. The phonon dispersions were calculated using a finite-displacement method using a $2 \times 2 \times 1$ supercell. Prior electronic structure calculations were carefully converged with denser k-point grids ($8 \times 8 \times 1$, $4 \times 4 \times 1$, as shown in Figure S1. We utilize Ti d-orbitals and N p-orbitals to construct the maximally localized Wannier functions (MLWF) [26].The DFT band structure within 2 eV from the Fermi level can be well-reproduced through the constructed Wannier functions. Finally, the Fermi-Golden Rule was implemented to calculate the inverse e–ph relaxation time [27]:

$$\left(\tau_{\mathrm{e-ph}}^{-1}\right)_{kn} = \frac{2\pi}{\hbar}\int_{\mathrm{BZ}}\frac{\Omega dk'}{(2\pi)^3}\sum_{n'\alpha\pm}\delta\left(\epsilon_{k'n'}-\epsilon_{kn}\mp\hbar\omega_{k'-k,\alpha}\right) \times\left(n_{k'-k,\alpha}+\frac{1}{2}\mp\left(\frac{1}{2}-f_{k'n'}\right)\right)\left|g_{k'n',kn}^{k'-k,\alpha}\right|^2$$

Where $\hbar\omega_{k\alpha}$ and $n_{k,\alpha}$ denote the energy and Bose occupation numbers of the phonon state, respectively, associated with wave vector and polarization index $\alpha$. Additionally, the term $g_{k'n',kn}^{k'-k,\alpha}$ represents the electron–phonon matrix element between the electron state k,n and k',n' via a phonon mode with a momentum k'-k. It is essential to note that the sum over ± effectively captures the dynamics of both phonon absorption and emission, highlighting the intricate interplay between these processes. To compute the e–ph lifetimes, we use $10^4$ Monte Carlo sampling points for the summation of the above equation. The delta function is approximated by a Gaussian with a small broadening value of 0.02 eV. The final lifetime is computed at the room temperature, where the Bose and Fermi occupation factors are applied to include the temperature. The scattering rate is governed by the e-ph matrix elements and energy-conserving phase space.

***2.2 Experimental details:*** To mimic the crystallographic orientations, we deposited TiN thin films on MgO substrates aligned with [100], [110], and [111] crystal orientations using radiofrequency magnetron sputtering [28, 29]. All measured TiN films were precisely deposited to a thickness of 100 nm under meticulously optimized conditions. In addition, TiN films with a thickness of 10 nm were prepared solely for cross-sectional TEM characterization. We selected MgO as the substrate due to its negligible lattice mismatch with TiN (0.7%), ensuring superior compatibility [30]. Table S1 depicts a comparison between the lattice constants of TiN and different substrates. During the sputtering process, high-purity argon was flowed into the chamber with a flow rate of 12 sccm. Detailed deposition conditions are listed in Table S2. To confirm the alignment of the TiN crystal planes with the MgO crystal orientations, we conducted X-ray diffraction for thorough characterization. Furthermore, in the ultrafast pump–probe transient absorption spectroscopy (TAS) measurements, we utilized a 400 nm pump laser, with a pulse duration of 100 fs and a repetition rate of 1 kHz, to excite hot carriers within the TiN film and employed continuous white light (450-700 nm) as the probe to investigate e-ph based hot carrier lifetimes [30]. The pump and probe spot sizes were approximately $7 \times 10^{-4}$ cm$^2$ and $1.6 \times 10^{-4}$ cm$^2$, respectively, ensuring that the probed area was fully contained within the uniformly excited region. All measurements were performed at room temperature, and the relative humidity was maintained between 40% and 50% in order to establish a controlled baseline comparison across different crystallographic orientations, while the relative anisotropic trends are expected to remain robust under identical thermal conditions.

To evaluate the e-ph relaxation lifetime, $\tau_1$, the transient absorption kinetics were analyzed using a global fitting procedure applied over the probe wavelength range of 520-550 nm. In this approach, the kinetic traces at different probe wavelengths were fitted simultaneously while sharing the same lifetime constants. To account for the multi-timescale carrier relaxation behavior, the fitting model includes an initial fast decay component, a rising and decaying intermediate component, and additional slower relaxation terms, as expressed below:

$$\Delta \mathrm{mod} = A e^{\frac{-t}{\tau_1}} + B\left(1 - e^{\frac{-t}{\tau_r}}\right) e^{\frac{-t}{\tau_2}} + C e^{\frac{-t}{\tau_3}} + D$$

Here, A, B, and C represent the amplitudes of the different dynamical components, $\tau_1$ corresponds to the ultrafast e-ph relaxation lifetime, $\tau_r$ describes the rise process, $\tau_2$ and $\tau_3$ denote the slower decay channels, and D is a constant offset. In general, the transient response of metallic materials involves e-e scattering, e-ph coupling, ph-ph coupling, and longer-timescale thermal effects [31, 32]. Here, we assign the fastest lifetime to e-ph coupling based on the two-temperature model (TTM) (see the details in the SI) [33]. Accordingly, electrons exchange energy with lattice phonons on a sub-picosecond timescale, which corresponds to the shortest lifetime observed in our measurements.

## 3. RESULTS AND DISCUSSION.

Figure 1 shows the slabs oriented in the [100], [110], and [111]-Ti directions, which are more than 2 nm thick to reduce the influence of quantum-well-states [34]. All slabs exhibit positive phonon frequencies, shown in Figure 2a–c, and positive zero-point energy (ZPE) values, shown in Table 1, suggesting that their structures are stable. Our stability trend of [111]-Ti > [100] > [110] > [111]-N is consistent with the earlier report [35]. For the [111] orientation, both Ti and N terminations are considered (Figure S3), and the Ti termination yield the lowest ZPE, as the more stable configuration.

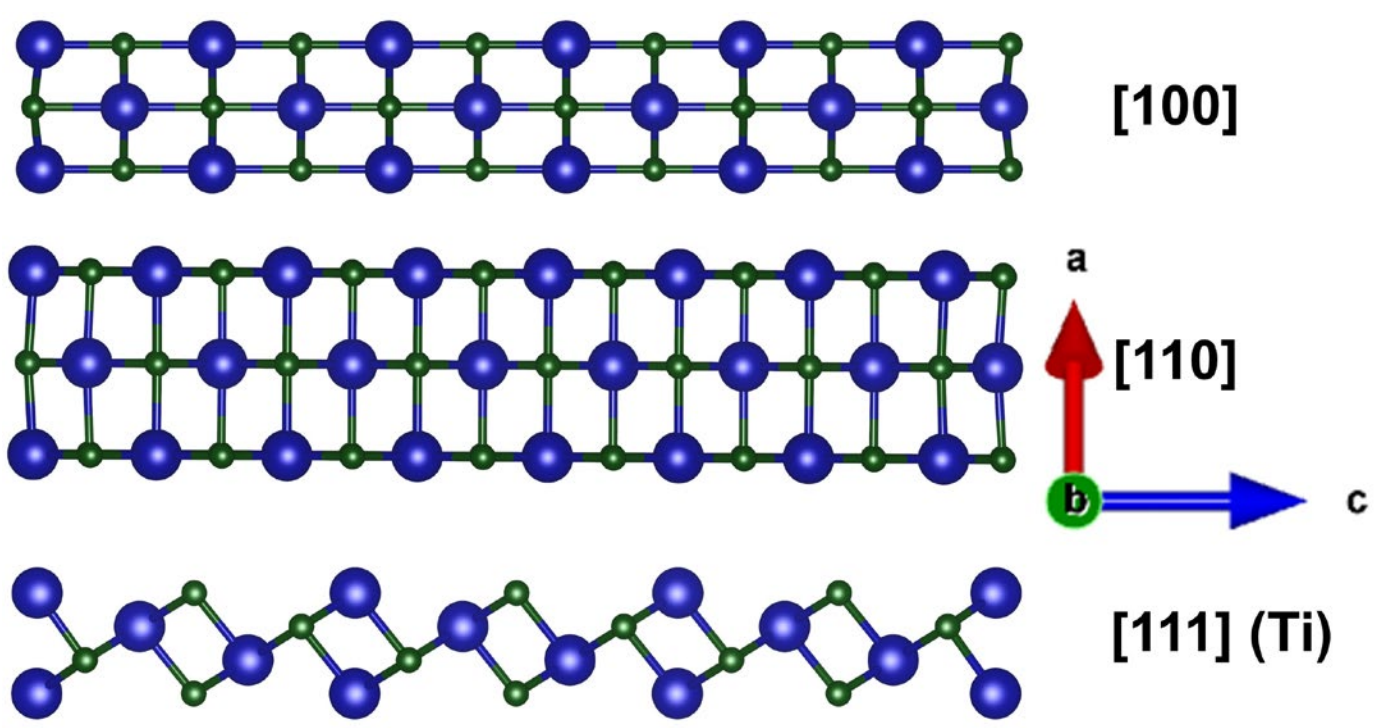

**Figure 1**. Optimized TiN slabs along [100], [110], and [111] (Ti) orientations, respectively. Blue (green) represents Ti (N) atoms.

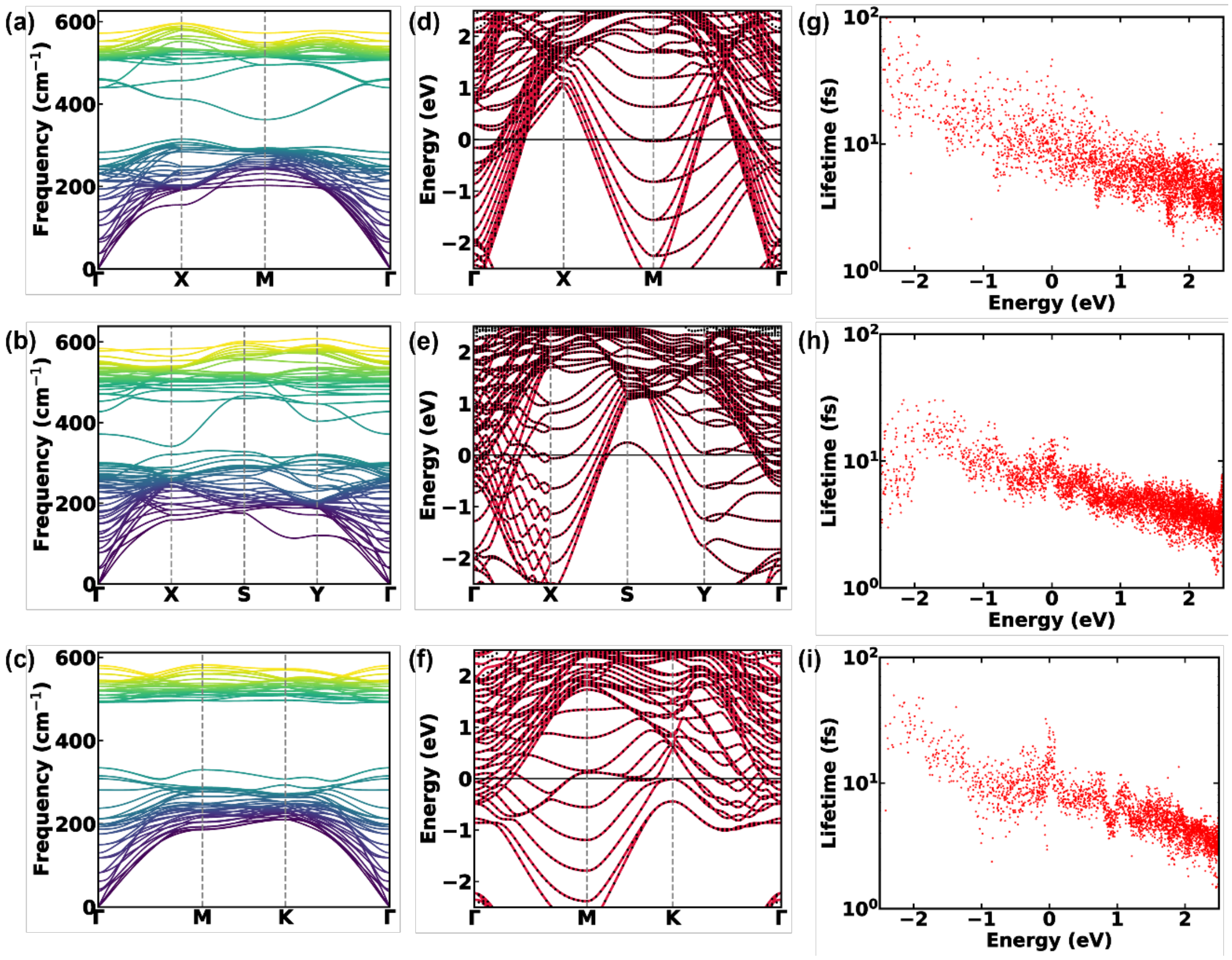

**Figure 2.** Calculated (a-c) slab phonons, (d-f) electronic DFT bands (black dotted) replotted with Wannier bands (red solid) and (g-i) e–ph scattering-based carrier lifetimes for [100], [110], and [111] (Ti) orientations, top to bottom respectively.

Using these slabs, we evaluate the orientation-specific hot-carrier lifetimes, defined as the inverse of the imaginary part of the electron-phonon self-energy, based on the phonon dispersions, electronic band structures, as given by the Eq. 1. Figure 2d-f and Figure S2 depicts the calculated electronic band structures and their orbital projection of the TiN slabs. We utilize MLWF to achieve a localized representation of these electronic band structures for the e–ph coupling calculations (see Methods). In addition, the band structures of the [111] orientation with N termination is shown in Figure S3.

We calculated the e–ph matrix element for the hot carrier lifetimes of TiN across assorted crystal orientations (see Methods), the hot carrier lifetimes for each orientation is determined by averaging within a 10 meV range around the Fermi level (Figure 2g-i) [36]. Conclusively, we obtain that the [111] orientation with a Ti termination has the longest carrier lifetime of 15.96 fs, followed by [100] and [110] orientations (13.96 and 11.12 fs, respectively), shown in Table 1. This longer lifetime in the [111] orientation originates from its reduced energy-conserving phase space for phonon-assisted scattering near the Fermi level, compared with those of the [100] and [110] orientations. We calculated the energy conservation phase space with respect to energy (Equation S1), indicating that how many phonon-assisted scattering pathways satisfy for phonon emission or absorption. A larger phase space corresponds to a shorter lifetime. The [110] orientation exhibits the largest phase space near and above the Fermi level (hot-electron region), followed by [100] and [111] orientations, shown in Figure S4, consistent with our lifetime results.

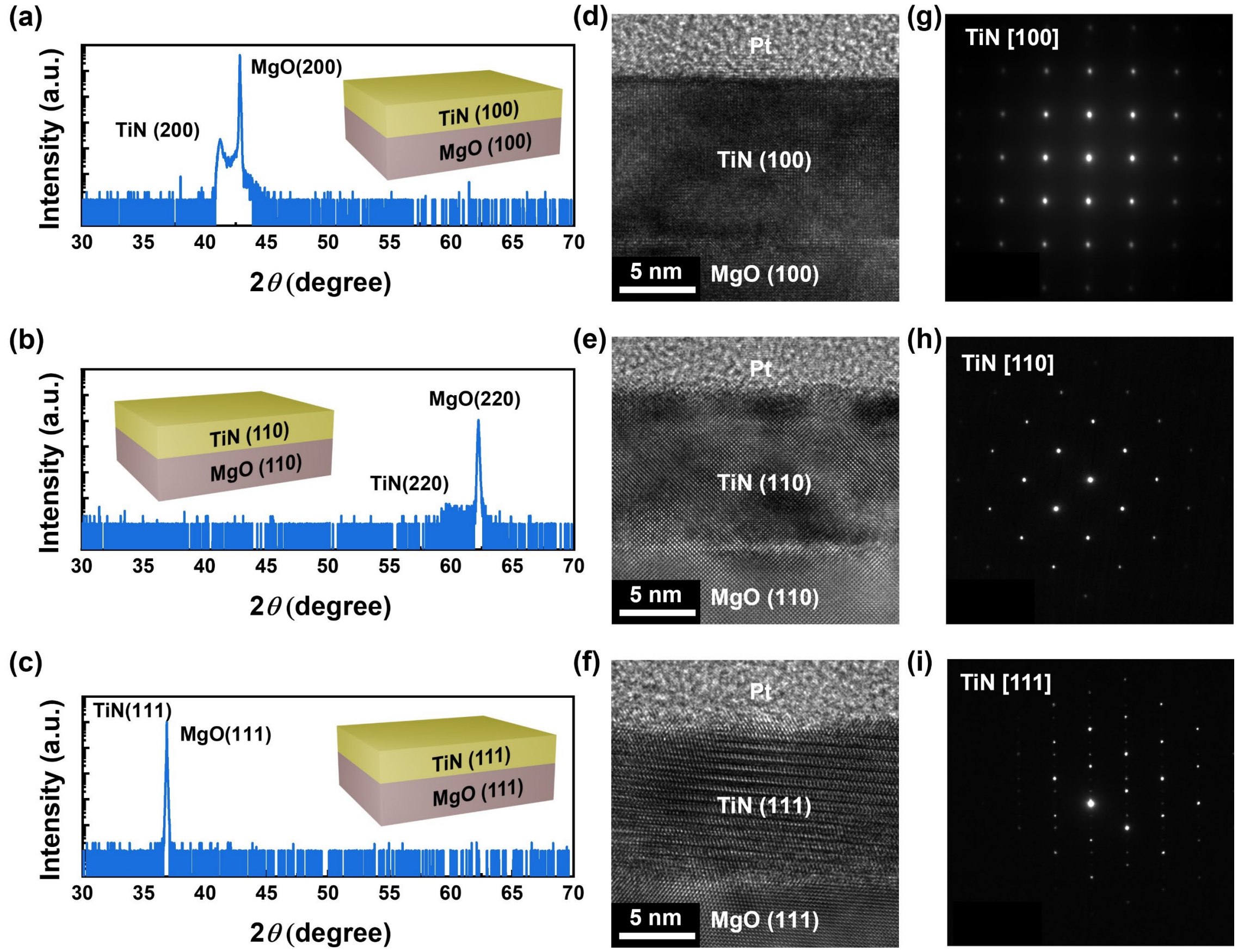

**Figure 3.** X-ray diffraction patterns of TiN along (a) [100], (b) [110], and (c) [111] crystallographic planes, showing the quasi-epitaxial growth of TiN. High-resolution cross-sectional transmission electron microscopy images show the quasi-epitaxial growth of TiN films on (d) [100], (e) [110], and (f) [111] MgO substrates. Selected area electron diffraction (SEAD) patterns obtained from TiN (g) [100], (h) [110], and (i) [111] orientations.

**Table 1:** Calculated zero-point energies (ZPE), and e–ph-based hot carrier lifetimes, and their measured values, for different crystal orientations.

| TiN Slab | ZPE (eV) | Cal. $\tau_{e\text{-}ph}$ (fs) | Exp. $\tau_{TAS}$ (fs) |
|---|---|---|---|
| **100** | 1.69 | 13.69 | 90 |
| **110** | 2.23 | 11.12 | 80 |
| **111** | 1.31 | 15.96 | 110 |

To validate the crystal orientations of the TiN thin films, we performed X-ray diffraction (XRD) and transmission electron microscopy (TEM) measurements, illustrated in Figure 3. In Figure 3a-c, the XRD peaks demonstrate that the crystal orientation of the TiN films directly mimics that of the MgO substrate for [100], [110], and [111] orientations, respectively. Additionally, the TEM images confirm the high quality of the TiN crystal orientations on the MgO substrates, as shown in Figure 3d-f. SEAD patterns obtained from TiN [100], [110], [111] orientations are shown in Figure 3g-i. In addition, we performed XPS depth profiling and chemical-state analysis of the quasi-epitaxial TiN (100), TiN (110), and TiN (111) films (Figures S5 and S6, and the details in the SI). The results confirm that oxygen impurities are confined to the as-received surfaces, where native oxides account for 44–51% of the Ti 2p signal, and become undetectable after sputtering (≤5%, the detection limit). The films possess a uniform nitrogen content, with 94–98% of the N 1s signal attributed to nitride species and comparable stoichiometries across the three orientations. No bulk oxidation is observed.

Figure 4a-c shows the TAS analysis of TiN thin films for the [100], [110], and [111] orientations, respectively. We also present UV-Visible absorption and transmission data, as shown in Figure S7. We presented various probe delay TAS in Figure S8, which offer a clearer insight into the temporal spectral evolution. Figure 4d represents the determined carrier lifetimes using global fitting (see experimental details) using the different probe wavelengths around 550 nm (shown in Table 1). The TiN [111] orientation exhibits the longest lifetime of 110 fs, followed by the [100] orientation of 90 fs and the [110] orientation of 80 fs. The lifetime decreases from the [111] to the [110] orientations align with our theoretical predicted trend. Our experimental findings indicate that crystal orientation provides a means to tune hot carrier lifetimes, driven by anisotropy in e-ph scatterings.

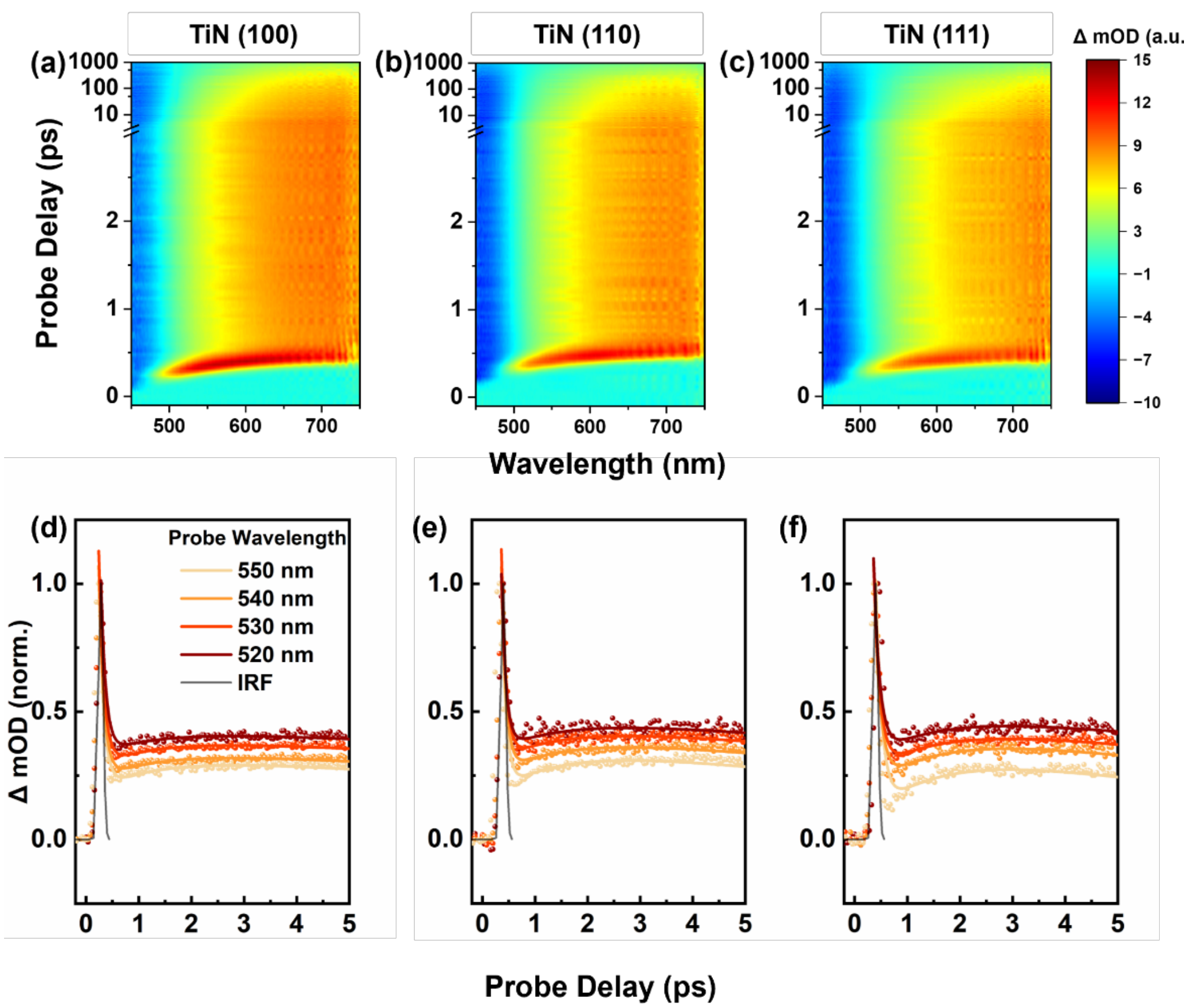

**Figure 4.** Color plots of time-resolved transient absorption spectroscopy (TAS) of TiN thin films along (a-c) [100], [110], and [111] orientations, respectively, (d-f) and their time-resolved TAS kinetics extracted at 520-550 nm and corresponding global fits for quasi-epitaxially grown TiN films with different orientations, highlighting e-ph lifetimes.

On the theory side, we compute the state-resolved e-ph quasiparticle lifetime, reflecting the underlying electronic structure and e-ph coupling strength. For experimental part, the pump-probe experiment tracks the macroscopic energy relaxation of the collective, non-equilibrium electron population [14, 40]. The measured decay can be further influenced by the available energy-conservation phase space for scattering. These timescales are therefore not expected to match one-to-one. Initially, the pump pulse energy is absorbed by TiN, a plasmonic metal, and creates a non-thermal electronic distribution (plasmon-assisted absorption) [37, 38]. These energetic electrons transfer their excess energy to the material's lattice (the phonons) with high efficiency. Our calculations show this electron-phonon scattering process happens on a femtosecond timescale. However, in the measurements, the subsequent dissipation of the energy from optical phonon (500–600 $cm^{-1}$) into acoustic phonons (under 300 $cm^{-1}$) modes can be slow due to their phonon bandgaps (Figure 2a-c) [39, 40]. This bottleneck leads to a massive, non-equilibrium buildup of hot optical phonons, where electrons can re-absorb energy from these phonons, slowing down the cooling rate [41-43]. To further understand the orientation-dependent cooling dynamics, Figure S9 shows the calculated electronic and lattice specific heats and e-ph lifetimes for the different orientations within the TTM analysis [43]. Notably, the [111] orientation shows the weakest e-ph coupling and the largest electronic specific heat, which together result in the longest lifetime (112.9 fs), followed by the [110] (83.9 fs) and [100] (78.6 fs) orientations. The measured lifetimes are 110, 90 and 80 fs for [111], [100] and [110] orientations, respectively. Since the TTM does not account for the energy-conserving phase space, the lifetime trend predicted is [100] < [110] < [111], whereas both the experimental results and first-principles calculations show the order of [110] < [100] < [111].

## 4. CONCLUSION.

In summary, we study the impact of [100], [110] and [111] orientations on e–ph scatterings and their effect on hot carrier relaxation dynamics in TiN thin films, combining first-principles calculations with pump-probe measurements. Our calculations predict that the Ti-terminated [111] orientation exhibits reduced e-ph scattering and yields the slowest relaxation, followed by [100] and [110] orientations. Our TAS measurements show that the pump-probe relaxation time, the collective cooling, increases in the order of [100], [110], and [111]. Our results indicate that the relaxation behaviors change with crystal orientations, and that can be useful depending on the device functions. For example, in the [111] TiN films the cooling is slower, so the hot-carriers stays around longer in a non-equilibrium form, relevant for photothermal conversion and refractory plasmonic uses where localized heating is the goal.

## ASSOCIATED CONTENT

The supplementary information (SI) is available free of charge. It includes the convergence of phonon dispersion, projected band structures for all orientations, and the electronic, phonon, and electron-phonon based lifetime for TiN (N-terminal).

## Credit authorship contribution Statement

**Hemant Verma:** Investigation, Writing – original draft, Data curation, Formal analysis, Visualization. **Tzu-Yu Peng:** Investigation, Writing – original draft, Data curation, Formal analysis, Visualization. **Shyr-Shyan Yeh:** Validation, Data curation. **Sheng-Chieh Huang:** Investigation, Validation. **Pritam Sardar:** Validation. **Yang-Hao Chan:** Supervision, Writing–review and editing. **Yu-Jung Lu:** Supervision, Writing–review and editing, Funding acquisition. **Chao-Cheng Kaun:** Supervision, Writing–review and editing, Funding acquisition.

## Acknowledgements

This work was financially supported by the Center for Advanced Semiconductor Technology Research under the Higher Education Sprout Project of the Ministry of Education, Taiwan, the National Science and Technology Council (Grant numbers NSTC 114-2112-M-001-059, NSTC 114-2811-M-001-144, NSTC 113-2112-M-001-014 and NSTC 114-2112-M-001-005), Academia

Sinica (ASTP-113-M02), and National Center for High-Performance Computing for providing computational and storage resources.